\documentclass[conference,letterpaper]{IEEEtran}

\usepackage{amsmath}
\usepackage{amssymb}
\usepackage{graphicx}
\DeclareMathOperator*{\argmax}{arg\,max}

\begin{document}

\title{Binary Rate Distortion With Side Information:\\ The Asymmetric Correlation Channel Case}
\author{\IEEEauthorblockN{Andrei Sechelea\IEEEauthorrefmark{1}, Samuel Cheng\IEEEauthorrefmark{2}, Adrian Munteanu\IEEEauthorrefmark{1}, and Nikos Deligiannis\IEEEauthorrefmark{1}}
\IEEEauthorblockA{\IEEEauthorrefmark{1}Department of Electronics and Informatics, Vrije Universiteit Brussel, Brussels, Belgium}\IEEEauthorblockA{\IEEEauthorrefmark{2}School of Electrical and Computer Engineering, University of Oklahoma, Tulsa, OK USA}\\
E-mail: \{atsechel, acmuntea, ndeligia\}@etro.vub.ac.be, samuel.cheng@ou.edu
}

\maketitle

\begin{abstract}Advancing over up-to-date information theoretic results that assume symmetric correlation models, in this work we consider the problem of lossy binary source coding with side information, where the correlation is expressed by a generic binary asymmetric channel. Specifically, we present an in-depth analysis of rate distortion with side information available to both the encoder and decoder (conventional predictive), as well as the Wyner-Ziv problem for this particular setup. Prompted by our recent results for the Z-channel correlation case, we evaluate the rate loss between the Wyner-Ziv and the conventional predictive coding, as a function of the parameters of the binary asymmetric correlation channel.
\end{abstract}


\section{Introduction}
Using side information is fundamental in many source and channel coding problems and has recently been adopted in new fields, like compressed sensing and compressive classification \cite{mota2014compressed}. The vast majority of results in discrete source coding with side information assume that correlation channel between the source and the side information to be symmetric. Side information is also a key factor in distributed source coding; e.g. recent results in distributed video coding \cite{Toto-Zarasoa12,DeligiannisTIP} have shown that using an asymmetric channel model to express the correlation between the source (frame to be encoded) and the side information (motion-compensated prediction)  leads to substantial compression performance gains compared to the case where a conventional symmetric correlation channel model is used. 

In this work, we focus on the theory of rate distortion with side information. Specifically, we consider the following setup: let $(X,Y)\in \mathcal{X} \times \mathcal{Y}$ be correlated binary random variables, such that the source $X\backsim\textit{Bernoulli}(0.5)$ is to be encoded using the source $Y$ as side information. The correlation between the two sources is described by a  binary asymmetric channel:
\begin{equation}
p(Y|X)=\begin{bmatrix} (1-a) & a \\ b & (1-b) \end{bmatrix},
\label{eq: correl_channel}
\end{equation}
$(a,b)\in\big[0,1\big]^2$. The reconstructed source is the binary variable $\hat X\in \hat{\mathcal{X}}$, and the distortion metric considered is the Hamming distance: if $(x,\hat x)\in \mathcal{X} \times \hat{\mathcal{X}}$ then $d_{H}(x,\hat x) = 0$ if $x=\hat x$, and $d_{H}(x,\hat x) = 1$ if $x \neq\hat x$. Our goal is to describe the rate distortion characteristics of this system in two cases, that is when the side information $Y$ is available at  (i) both the encoder and decoder, corresponding to the conventional predictive case, and (ii)  available only at the decoder,  i.e., the  Wyner-Ziv (WZ) case. 

For the binary symmetric correlation, the rate distortion functions were given by Berger  in \cite{berger1971rate} for the predictive case, and respectively by Wyner and Ziv in \cite{WynerZiv}, for the corresponding WZ case. Another binary correlation channel that has been recently studied is the Z-channel, for which Steinberg derived in \cite{steinberg2009coding} the rate distortion bound $R_{X|Y}^Z(d)$ for the predictive case, while in \cite{deligiannis2014no} we derived the rate distortion bound $R_{WZ}^Z(d)$ for the Wyner-Ziv case. In \cite{WynerZiv}, it is established that, for the general case, a rate loss exists between the predictive and WZ bounds. Zamir showed in \cite{zamir1996rate} that, for binary sources and the Hamming distortion metric, the rate loss is at most 0.22 bits/sample,  and in \cite{deligiannis2014no} we show that this difference vanishes in the case of the Z-channel correlation.

To the best of our knowledge, the characterization of the binary rate distortion function in the general case of  the asymmetric correlation is still an open problem. A straightforward option is to use a numerical algorithm to compute it, such as the Blahut-Arimoto \cite{blahut1972computation} implementation in \cite{cheng2005computing}. Unfortunately,  such a numerical analysis does not provide an  analytical characterization of the rate distortion function nor an understanding of its   behaviour in various rate regions.    

In this paper, we present the derivation of the conventional predictive $R_{X|Y}(d)$ function for the considered setup. Subsequently, by considering a binary auxiliary variable in the formulations of \cite{WynerZiv}, we derive rate and distortion expressions that yield a bound for the WZ problem.  Furthermore, through our analysis, we are able to evaluate the evolution of the rate-loss, with respect to the crossover probabilities of the binary asymmetric correlation channel.
 
\section{Source coding with Encoder-Decoder Side Information}
When the side information
is available both at the encoder and the decoder, the rate-distortion function is given by
\begin{equation}
R_{X|Y}(d)=\inf_{p(\hat x|x,y): \ E\big[d(X,\hat X)\big]\leq d} I(X;\hat X|Y),
\label{eq:RD_SI_ED}
\end{equation}
where $E\big[\cdot\big]$ denotes the expectation operator and $d$ is the targeted distortion. 

The quantity to be minimized in (\ref{eq:RD_SI_ED}) can be written as
\begin{align}
I(X;\hat X|Y) &= \ H(X|Y)-H(X|\hat X,Y) \nonumber \\
&= \ H( X|Y) - H(X \oplus \hat X|\hat X,Y) \nonumber \\
&\geq H(X|Y) - H(\mathcal{D}|Y),
\label{eq:RD_SI_ED_1}
\end{align}
where $H(\cdot )$ denotes the binary entropy function and $\mathcal{D}=X \oplus \hat X$. With the $X-Y$ correlation channel given by (\ref {eq: correl_channel}), the  maximum distortion to be considered is the average crossover of the side information channel, i.e., if $\hat X = Y$: $D_{max} = \frac{a+b}{2}$. We consider the inverse channel $Y-X$:
\begin{align}
p(X|Y)&=\begin{bmatrix} 1-a^*  & a^*  \\ b^* & 1-b^* \end{bmatrix}
\
&=\begin{bmatrix} \frac{(1-a)}{1-a+b} & \frac{b}{1-a+b} \\ \frac{a}{a+ 1-b} & \frac{(1-b)}{a+1-b} \end{bmatrix}.
\label{eq: inv_correl_channel}
\end{align}
   We have $p(Y=0) = \frac{1-a+b}{2}$ and $p(Y=1) = \frac{a+1-b}{2}$, and the maximum distortion can be written as $D_{max} = p(Y=0)\cdot a^*+p(Y=1)\cdot b^*$. Without loss of generality, let $a\le b$ in (\ref {eq: correl_channel}). This implies $a^*\ge b^*$.
Given that the side information is known at the encoder, we can always describe the overall distortion as being 
\begin{equation}
d = E\big[\mathcal{D}\big]=p(Y=0)\cdot d_{Y=0}+p(Y=1)\cdot d_{Y=1},
\end{equation}
where $d_{Y=0} =E\big[d(X, \hat X)|Y=0\big] \leq a^*$ and $d_{Y=1} =E\big[d(X, \hat X)|Y=1\big] \leq b^*$. 

Maximizing the term $H(\mathcal{D}|Y)$ in (\ref{eq:RD_SI_ED_1}) with the above mentioned constraints gives the minimum for $I(X;\hat X|Y)$. This is an optimization problem that can be expressed using the Karush-Kuhn-Tucker conditions:
\begin{align}
&\text{maximize} \quad & p(Y=0)\cdot H(d_{Y=0})+p(Y=1)\cdot H(d_{Y=1})  \nonumber\\
&\text{subject to} \quad &
\begin{cases}
p(Y=0) \cdot d_{Y=0} + p(Y=1) \cdot d_{Y=1}= d \\
d_{Y=0}\le a^* \\
d_{Y=1}\le b^*
\end{cases}
\label{eq:KKTpb}
\end{align}
We formulate the Lagrangian optimization problem with the above inequality constraints, and the Lagrangian function is:
\begin{equation}
\begin{split}
\mathcal{J} = p(Y=0)\cdot H(d_{Y=0})+p(Y=1)\cdot H(d_{Y=1}) +  \\
\lambda(p(Y=0) \cdot d_{Y=0} + p(Y=1) \cdot d_{Y=1}-d) + \\
\lambda_{0}(d_{Y=0}-a^*)+\lambda_{1}(d_{Y=1}-b^*)  
\end{split}
\end{equation}
 Deriving with respect to the unknowns yields:
\begin{equation}
R_{X|Y}(d) = 
\begin{cases}
p(Y=0)\cdot \big[H(a^*)-H(d)\big]+ \\ \qquad p(Y=1)\cdot\big[H(b^*)-H(d)\big], \ \ \mbox {if} \ d\le b^* \\ 
p(Y=0)\cdot \big[H(a^*)-H(\frac{d-\frac {a}{2}}{p(Y=0)})\big], \\ \qquad \ \ \  \mbox {if} \  b^*\le d \le D_{max} \\ 
0, \qquad \mbox{if} \ d\ge D_{max}
\end{cases}
\label{eq:R_X|Y_D}
\end{equation}
This can be achieved by considering an auxiliary variable $U$ given by a binary symmetric channel with input $X$ and output $U$, with crossover probability $p_0$ as follows:
\begin{equation}
p_0=\begin{cases}
\mbox {if} \ d\le b^* & \mbox{then} \ p_0 = d  \\
\mbox {if} \ d> b^* & \mbox{then} \ p_0 = \frac{d-p(Y=1)\cdot b^*}{p(Y=0)}
\end{cases}
\end{equation}
The corresponding reconstruction function is:
\begin{equation}
\begin{cases}
\mbox {if} \ d\le b^* & \mbox{then} \ \hat X = U  \\
\mbox {if} \ d> b^* & \mbox{then} \ 
\begin{cases}
\hat X = Y & \mbox{if} \ Y = 1 \\
\hat X = U & \mbox{if} \ Y = 0 
\end{cases}
\end{cases}
\end{equation}
\section{Source Coding with Side Information at the Decoder}
When the side information is available only at the decoder, the R-D function is given by
\begin{equation}
R_{WZ}(d)=\inf_{p(u|x)p(\hat x|u,y): \ E[d(X,\hat X)]\leq d} I(X;U|Y),
\label{eq:RD_SI}
\end{equation}
where $\hat{X}=f(U,Y)$ and $U$ is an auxiliary random variable, satisfying the Markov chains: $U\leftrightarrow X\leftrightarrow Y$ and $X\leftrightarrow (U, Y)\leftrightarrow \hat X$, such that $E\big[d(X,\hat X)\big]\leq d$.
Let $U \in \mathcal{U}$ be the outcome obtained from a binary channel with input $X$ and the transition matrix given by:
\begin{equation}
p(U|X)=\begin{bmatrix} (1-p) & p \\ q & (1-q) \end{bmatrix},  
\label{eq:xu_channel}
\end{equation}
with $(p,q)\in\big[0,1\big]^2$.

We will  express  the resulting rate and distortion as functions of the crossover probabilities of the channels in (\ref{eq: correl_channel}) and (\ref{eq:xu_channel}).

\subsection{Expression of the Rate}
The quantity to be minimized in (\ref{eq:RD_SI}) is written as
\begin{equation}
 I(X;U|Y) = H(U|Y)-H(U|X).
 \label{eq:WZ_exp}
\end{equation}
Given the Markovianity $U\leftrightarrow X\leftrightarrow Y$ the channel between $Y$ and $U$ can be expressed as the concatenation of the channels $X-U$ and $X-Y$. Knowing (\ref{eq: inv_correl_channel}) and (\ref{eq:xu_channel}) the transition matrix is 
\begin{align}
p(U|Y) 
  &=  \begin{bmatrix} \frac{(1-a)(1-p)+ bq}{1-a+b} & \frac{(1-a)p+ b(1-q)}{1-a+b} \\ \frac{a(1-p)+(1-b)q}{a+1-b} & \frac{ap+(1-b)(1-q)}{a+1-b} \end{bmatrix}.
\end{align}
Therefore, based on (\ref{eq:WZ_exp}), let us define $R_{WZ}^*(p,q)$ to be:
\begin{align}
I(X;U|Y)&\triangleq\ R_{WZ}^{*}(p,q) = H(U|Y)-H(U|X)   \nonumber
\\
&= \frac{(1-a+b)}{2} \cdot H\left(\frac{(1-a)(1-p)+ bq}{1-a+b}\right) \nonumber\\ & + \frac{(a+1-b)}{2} \cdot H\left(\frac{a(1-p)+(1-b)q}{a+1-b}\right)  \nonumber\\
&\qquad\qquad\qquad-\frac{1}{2}\cdot\big[H(p)  + \  H(q)\big].
 \label{eq:Rate_general}
 \end{align}

\subsection{Expression of the Distortion}
The distortion expression can be determined by observing that for a fixed pair $(u,y)$, $x$ is given by the conditional distribution $p(x|u,y)$. Since the decoder can only make a deterministic decision given $u$ and $y$, the best choice is to output:
\begin{equation}
 \hat x = f(u,y) = \argmax_x p(x|u,y). 
 \label{eq:reconstruction_function}
\end{equation}
We can write the error rate conditioned on $(u,y)$ as $\big[1-p(f(u,y)|u,y)\big]$, so the average distortion is given by
\begin{equation}
\label{eq:Dist} 
d = \sum_{u,y} (1-p(f(u,y)|u,y) ) p(u,y).
\end{equation}
This is equivalent to:
\begin{equation}
\label{eq:Dist_qp} 
d = \sum_{u,y} \min( \ p(X=0,y,u),\ p(X=1,y,u) \ ).
\end{equation}
Knowing that, due to the Markov property $U\leftrightarrow X\leftrightarrow Y$, 
\begin{equation}
p(x,y,u) = p(x,y)p(u|x)=p(x)p(y|x)p(u|x)
\end{equation}
 we can write $p(x,y,u) $ as follows:\begin{equation}
\label{eq:X0YU}
p(X=0,Y,U)=\frac{1}{2} \cdot \begin{bmatrix}(1-a)(1-p) & (1-a)p \\  a(1-p) & ap \end{bmatrix},
\end{equation}
and
\begin{equation}
\label{eq:X1YU}
p(X=1,Y,U)=\frac{1}{2} \cdot  \begin{bmatrix} bq & b(1-q) \\ (1-b)q & (1-b)(1-q) \end{bmatrix}.
\end{equation}

At this point, it is useful to make the following observations regarding the symmetries of the $R_{WZ}^{*}(p,q)$ and $d$ with respect to the crossover probabilities in (\ref{eq: correl_channel}) and (\ref{eq:xu_channel}):
\begin{itemize}
\item If in (\ref{eq: correl_channel}) we substitute the pair $(a,b)$ by $(1-a,1-b)$, the functions $R_{WZ}^{*}(p,q)$ and $d$ do not change.
\item If in (\ref{eq:xu_channel}) we substitute the pair $(p,q)$ by $(1-p,1-q)$, the functions $R_{WZ}^{*}(p,q)$ and $d$ do not change.
\end{itemize}
Essentially, the substitutions are equivalent to a label swap at the output of the respective channels and do not affect the rate-distortion function. The consequence is that the domains of interest for the pairs $(a,b)$ and $(p,q)$ can be reduced from $[0,1]^2$ to $0\leq a+b \leq 1$ and $0\leq p+q\leq 1$. Using symmetry constraints and (\ref{eq:X0YU}), (\ref{eq:X1YU}),     equation (\ref{eq:Dist_qp}) becomes:
\begin{align}
\mathcal{D}(p,q) = &(bq + \min((1-a)p,b(1-q))+ \nonumber \\ 
& \min(a(1-p),(1-b)q)+ap)/2. 
\label{eq:Dist_4_cases}
\end{align}
There can be four possible expressions for the above equation, each corresponding to a different reconstruction strategy at the decoder.
Without loss of generality, let $a<b$, fixed by the initial setup. The crossover probabilities $p$ and $q$ can vary. 

In order to emphasize the decision taken by the reconstruction function given in (\ref{eq:reconstruction_function}) for every pair $(y,u)$, we introduce the following function:
\begin{equation}
\hat x_{out}(y,u) = \argmax_x p(x, Y=y,U = u)
\end{equation}
In our binary case, $\hat x_{out}$ can be seen a $2 \times 2$ matrix, with the lines corresponding to the values of $Y\in\{0,1\}$, and columns corresponding to the values of $U\in \{0,1\}$,  indicating whether the reconstruction was $\hat x = 0$ or $\hat x = 1$.  

The possible expressions for (\ref{eq:Dist_4_cases}) are the following:
\begin{itemize}
\begin{item}
when $ \begin{cases}
(1-a)p< b(1-q) \\
a(1-p)>(1-b)q
\end{cases},$
$\hat{x}_{out} = \left (\begin{array}{cc}0 & 1 \\ 0 &1 \end{array} \right)$ is equivalent to having the reconstruction function $\hat{X} = U$ and a distortion value of $D_1 = \frac{p+q}{2}$;
\end{item}
\begin{item}
when $\begin{cases}
(1-a)p< b(1-q) \\
a(1-p)<(1-b)q
\end{cases},$
$\hat{x}_{out} = \left (\begin{array}{cc}0 & 1 \\ 1 &1 \end{array} \right)$ is equivalent to having the reconstruction function $\hat{X} = Y \lor U$ (where $\lor$ is the binary OR operator) and a distortion value of $D_2 = \frac{bq+(1-a)p+a}{2}$;
\end{item}
\begin{item}
when $\begin{cases}
(1-a)p> b(1-q) \\
a(1-p)>(1-b)q
\end{cases},$
$\hat{x}_{out} = \left (\begin{array}{cc}0 & 0 \\ 0 &1 \end{array} \right)$ is equivalent to having the reconstruction function $\hat{X} = Y \land U$ (where $\land$ is the binary AND operator) and a distortion value of $D_3 = \frac{(1-b)q+ap+b}{2}$;
\end{item}
\begin{item}
when $\begin{cases}
(1-a)p>b(1-q) \\
a(1-p)<(1-b)q
\end{cases},$
$\hat{x}_{out} = \left (\begin{array}{cc}0 & 0 \\ 1 & 1 \end{array} \right)$ is equivalent to having the reconstruction function $\hat{X} = Y$ and a distortion value of $D_4 = \frac{a+b}{2}$.
\end{item}
\end{itemize}
 
\begin{figure}
\center
\includegraphics [height=1.7in]{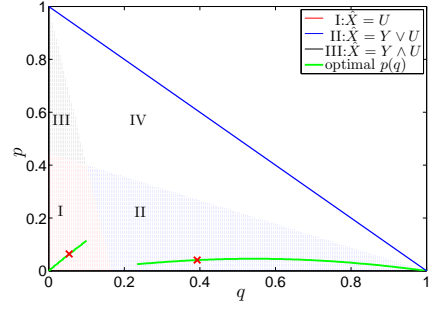}
\caption{Reconstruction regions in the $(p,q)$ plane for $a = 0.1, b = 0.4$}
\label{fig:pq_1}
\end{figure}

An example of the resulting decision regions derived above is presented in Fig. \ref {fig:pq_1}. The white region (IV) corresponds to the reconstruction decision $\hat{X}=Y$. When the pair $(a,b)$ is fixed, the lines that delimit the regions in the plot are given by: $a(1-p)=(1-b)q$ and $(1-a)p = b(1-q)$. 

\subsection{Deriving a tight bound on the rate distortion function}
         
Given $R_{WZ}^{*}(p,q)$ as in (\ref{eq:Rate_general}) and the four expressions of the distortion function in (\ref{eq:Dist_4_cases}), finding the bound can be formulated as a minimization problem with constraints. For every distortion level $d \in \big[0, D_{max}\big]$,  the goal is to find the $(p,q)$ pair that minimizes the rate, while satisfying the distortion constraint:
\begin{align}
& \text{for each $d \in\big[0,D_{max}\big]$ minimize} \ R_{WZ}^*(p,q) \nonumber \\
& \text{subject to}  
\begin{cases}
0\le \mathcal{D}(p,q)\le d\le D_{max}\\
\mathcal{D}(p,q) \ \text{has a form} \in \{ D_1, D_2,D_3,D_4\}
\end{cases}
\label{eq:min_pb}
\end{align}
The result of the above minimization is denoted $R_{WZ}^*(d)$.

In order to establish a characterization of the bound, we need to show how a certain distortion level can be achieved, i.e., what reconstruction function should be used. For example, if the required distortion level is $d=0$, this can only be obtained using $\hat X = U$, since $\min(D_2) = a/2$, $\min(D_3) = b/2$, and $D_4$ is a constant.

As such, we consider all possible values for the expected distortion $d$,  and we must find what pairs $(p,q)$ can achieve it and the corresponding reconstruction strategy. As $d$ grows from zero to $D_{max}$, we can derive the following conclusions:
\begin{itemize}

\item  If $0\le d<\frac {a}{2 \cdot(1-b)}$ we have only one reconstruction possibility, denoted $RI$, namely $\hat X = U,$ and $d = D_1$.
\item If $\frac {a}{2 \cdot(1-b)} \le d \le \frac {b}{2 \cdot(1-a)}$ we have two possible reconstructions,  $RI$ and $RII$ (i.e. $\hat X = U \lor Y$), and $d$ can have forms $D_1$ or $D_2$.
\item If $\frac{b}{2 \cdot (1-a)} \le d \le D_{max}=\frac {a+b}{2}$ we have three possible reconstructions, cases $RI$, $RII$ and $RIII$ (i.e. $\hat X = U \land Y$), so $d$ can have forms  $D_1, D_2$ or $D_3$.
\item The reconstruction function $\hat X=Y$ gives constant distortion $D_4 =D_{max}$. \end{itemize}
\begin{figure}
\center
\includegraphics [height=1.7in]{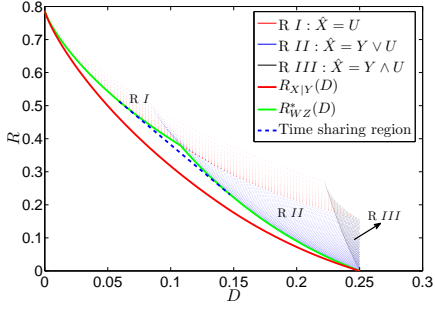}
\caption{$R_{X|Y}(d)$ and $R_{bound}(d)$ for a = 0.1 and b = 0.4}
\label{fig:RD_all_1}
\end{figure}
By numerical evaluation, the minimum for $R_{WZ}^*(d)$ can be achieved by letting $\hat X=U$ at lower distortion levels, and by letting $\hat X = U \lor Y$ for high distortions. 

We want our bound to be  a convex function; $R_{WZ}^*(d)$ on the other hand may not be convex, as it is the union of two convex curves: one corresponding to  $\hat X=U$, the other to $\hat X = U \lor Y$. To this end, we must consider the lower convex envelope of the achievable points: $R_{bound}(d) = \text{l.c.e.} \{ R_{WZ}^*(d)\}$. This will include the common tangent of the two curves and the region is achievable through time sharing. 

As an illustration, the rate distortion points in region $Ri$ of Fig. \ref{fig:RD_all_1} correspond to the $p(q)$ values in region $i$ of Fig. \ref{fig:pq_1}, for all $i\in \{I ,II, III\}$.  Fig. \ref{fig:RD_all_1} also presents $R_{WZ}^*(d)$, the common tangent which completes $R_{bound}(d)$, and the predictive $R_{X|Y}(d)$. 

The solution to the minimization problem proposed has a complex form and the resulting equations cannot be solved analytically. We can therefore obtain the $R_{bound}(d)$ curve only by numerically solving a logarithmic equation.

An important aspect is to answer whether the obtained rate distortion bound is tight.
From \cite{WynerZiv} we know that the cardinality of the auxiliary variable $U$ is bounded by $|\mathcal{X}| +1$, so a ternary alphabet for $U$ might be needed to achieve the rate distortion function. However, numerical experiments obtained with various probability input distributions $\pi$, and crossover probabilities of the correlation channel $(a,b)$, indicate that the proposed rate distortion bound does overlap with the  rate distortion function obtained with the Blahut-Arimoto algorithm in \cite{cheng2005computing}. Hence, we conjecture that a binary auxiliary variable $U$ is sufficient to achieve the rate distortion function and the proposed rate distortion bound $R_{bound}(d)$ is tight. A thorough analysis of this aspect is left as topic of further investigation.
\section {Optimal $p(U|X)$ and rate-loss assessment}
Since our bound $R_{bound}(d)$  does not have an analytical form, we use a numerical approach to derive the optimal crossover values of the $X-U$ channel,  i.e., the $(p,q)$ pairs, and also to find the time-sharing region.

\begin{figure}
\center
\includegraphics [height=1.1in,width=2in]{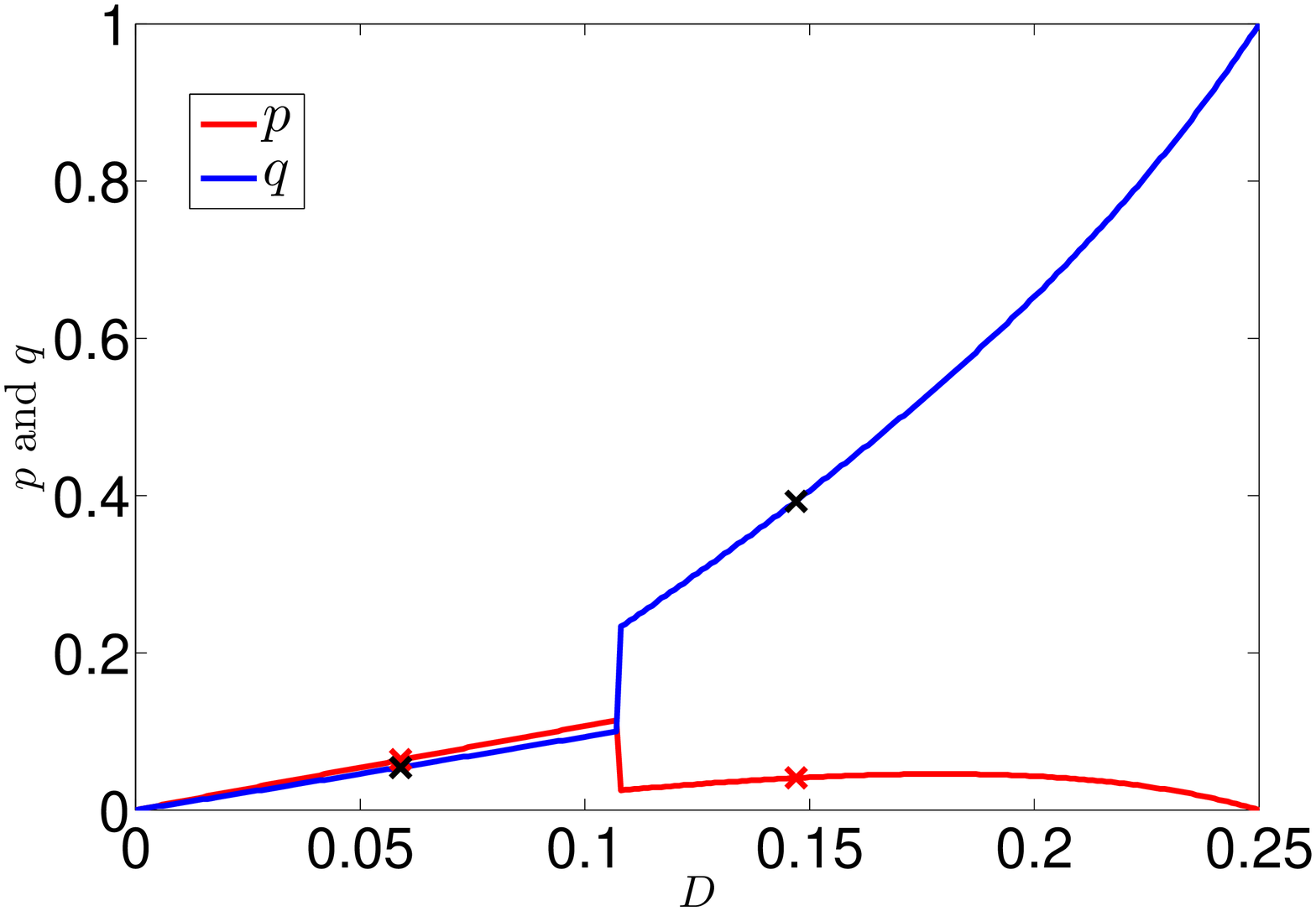}
\caption{Optimal $p(d)$ and $q(d)$ for $a = 0.1$ and $b = 0.4$}
\label {fig:pq_of_D}
\end{figure}

The optimal crossover pairs $(p,q)$ that achieve the $R_{WZ}^*(d)$ bound, are plotted as $p(q)$ in Fig. \ref{fig:pq_1}, while Fig. \ref{fig:pq_of_D} presents the same $(p,q)$ pairs  as functions of the distortion. The cross markers in both figures show the values that delimit the time sharing region. For low distortions, when the reconstruction function is $\hat X = U$, the optimal channels are close to binary symmetric channels. When the transition to the reconstruction function $\hat X = U \lor Y$ occurs, there is a discontinuity in the $p(d)$ and $q(d)$ functions. That can be noticed in Fig. \ref {fig:pq_1} as well, where the two green curve segments are disjoint. 

\begin{figure}
\center
\includegraphics [height=1.1in, width = 2in]{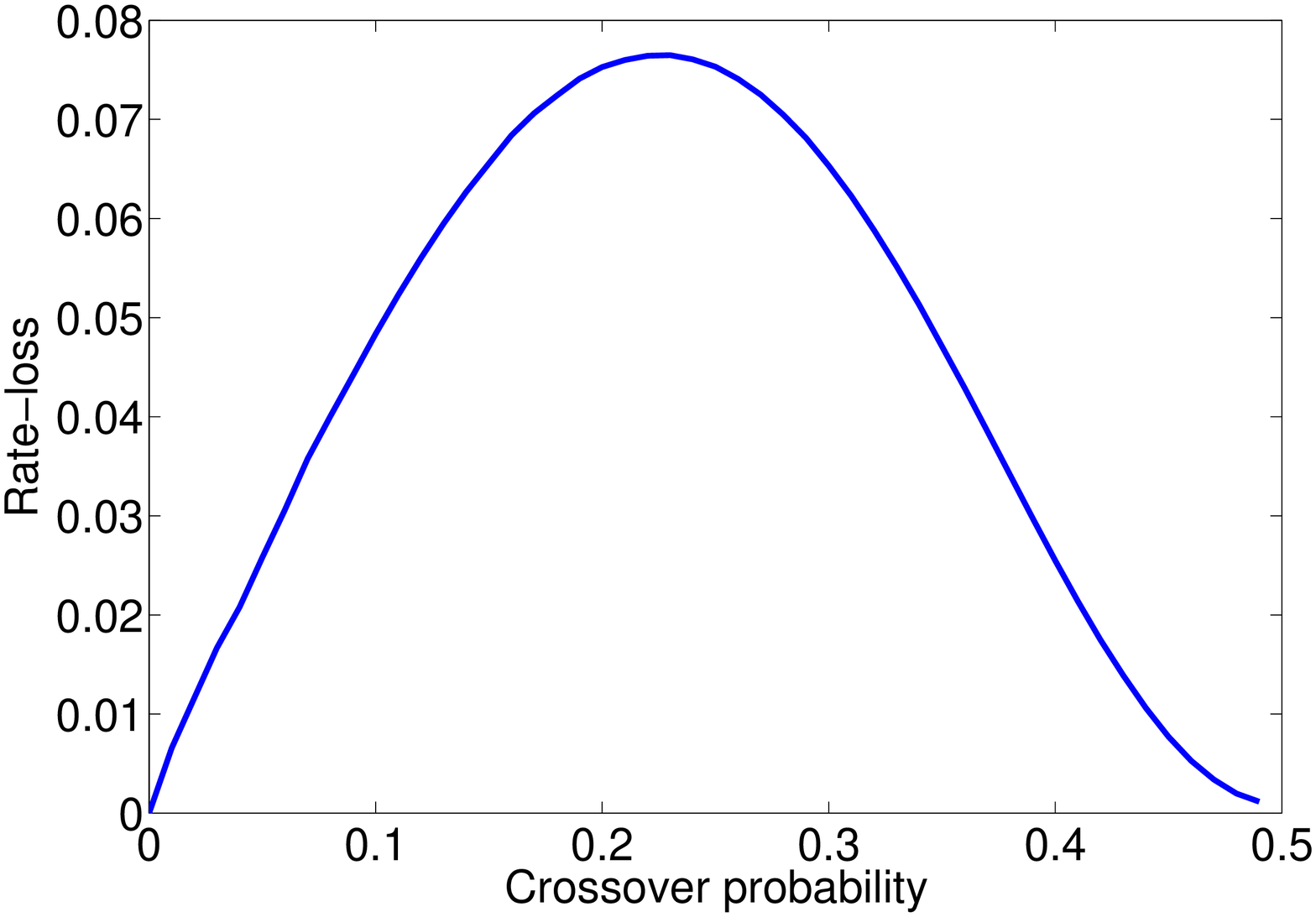}
\caption{Rate-loss for BSC correlation  with different crossover probabilities }
\label {fig:rateloss}
\end{figure}
If we fix the average distortion of the correlation channel, the highest rate-loss is in the case of the binary symmetric correlation. Fig. \ref{fig:rateloss} shows the variation of the rate-loss for different crossover probabilities of the binary symmetric correlation channel $X-Y$. The maximum value $\Delta R = 0.0765$ is obtained for the crossover $a = b = 0.227$. 

\section{Conclusion}

This paper analyzes   the rate distortion problem for a binary uniform source in the presence of correlated side information, when the correlation is given by an asymmetric channel. We have derived the $R_{X|Y}(d)$ for the conventional predictive case, and also proposed a bound $R_{bound}(d)$ for the Wyner-Ziv case. Numerical algorithms allow us to describe the optimum $p(U|X)$ achieving the bound, as well as to establish the maximum rate-loss in the case of binary uniform sources.


\bibliographystyle{IEEEtran}
\bibliography{references}

%

\end{document}